\documentclass[preprint]{aastex}
\usepackage{inputenc}
\usepackage[T2A]{fontenc}
\usepackage[english]{babel}
\usepackage{amsmath}
\usepackage{amssymb}
\usepackage{bm}
\usepackage{color}

\newcommand{\vecb}[1]{{\bm{\mathrm{#1}}}}

\begin{document}

\title{The Influence of the Decay of OB Associations on the Evolution of Dwarf Galaxies}

\author{E. P. Kurbatov}
\affil{Institute of astronomy, Russian Academy of Sciences}
\affil{48 Pyatnitskaya st., Moscow, Russian Federation, 119017}
\email{kurbatov@inasan.ru}


\begin{abstract}
It is commonly believed that most of the stars born in associations decaying
with characteristic velocities of stars $\sim 10$~km$/$s. For dwarf galaxies
the decay can lead to ejection of stars from the galaxy. The effect is studied
for spheroidal and disk dwarf galaxies, and is shown to have substantional
observational consequences for disk galaxies with escape velocities up to
$20$~km$/$s, or dynamical masses up to $10^8 M_\odot$. The ejection of stars
can (i) reduce the abundances of the products of Type Ia supernovae and, to a
lesser degree, Type II supernovae, in disk stars, (ii) chemically enrich the
galactic halo and intergalactic medium, (iii) lead to the loss of $50\%$ of the
stellar mass in galaxies with masses $\sim 10^7 M_\odot$ and the loss of all
stars in systems with masses $10^5 M_\odot$, (iv) increase the
mass-to-luminosity ratio of the galaxies.
\end{abstract}

\section{INTRODUCTION}

Mass exchange between a galaxy and the intergalactic medium(IGM) can influence
both the chemical composition of the galactic gas and the morphology of the
galaxy. Several mechanisms for gas loss by a galaxy can be distinguished
\citep{Shustov--1997A&A...317..397S}: galactic wind induced by numerous
supernova explosions, ram pressure exerted by the IGM, the tidal influence of
other galaxies of the group, evaporation of gas due to interactions with the
hot IGM, and blowing of the gas out of the galaxy by stellar
radiation. Possible ways for the loss of stellar mass include tidal
interactions, the ejection of stars due to the statistical mechanism, and the
decay of stellar associations. Let us consider some of these mechanisms in more
detail.

There have been many studies of the effects of numerous supernova explosions,
such as galactic fountains, super-bubbles, and winds, on disk galaxies
(see \citet{Shustov--1997A&A...317..397S,Cooper--2008ApJ...674..157C} and
references therein). The efficiency of these processes for gas ejection depends
strongly on the distribution of the gas. Thus, models of galaxies with a
stratified interstellar medium (ISM) display a much higher mass-loss efficiency
than do models with a continuous distribution of their ISM
\citep{Cooper--2008ApJ...674..157C}. Computations of models with a continuous
gas distribution, in turn, provide different results depending on the
distribution law \citep{Mac_Low--1989ApJ...337..141M}. It was shown in the
theoretical study of \citet{De_Young--1990ApJ...356L..15D} that the fraction of
expelled gas in a $1.4 \times 10^9~M_\odot$ galaxy is $\sim 0.6$, but, as the
authors note, the presence of dark matter was not taken into account. According
to \citet{Igumenshchev--1990A&A...234..396I}, galaxies with masses exceeding
$10^{12}~M_\odot$ do not have winds, and, hence, do not lose gas via this
mechanism. Based on these computations, \citet{Shustov--1997A&A...317..397S}
used a simple approximation for the relation between the mass of the galaxy and
the fraction of expelled matter in their models:
\begin{equation}
  \label{eq:galactic-wind}
  f_\mathrm{esc} = 2.4 - 0.2 \lg\frac{M_\mathrm{G}}{M_\odot}  \;.
\end{equation}
In this approximation, the efficiency of gas ejection becomes unity for
galaxies with masses of $10^7~M_\odot$, irrespective of their
morphology. According to this model, galaxies with such masses should not
contain gas. At the same time, gas is almost completely absent only in
spheroidal and elliptical dwarf galaxies, whereas it can constitute a
substantial fraction of the masses of disk and irregular galaxies
\citep{Begum--2004A&A...413..525B,Karachentsev--2004AJ....127.2031K,
Begum--2008MNRAS.383..809B}. On the other hand, observations have not revealed
gas outflows into the IGM from galaxies with dynamical masses of
$\sim 10^9~M_\odot$ \citep{van_Eymeren--2009A&A...493..511V}. Thus, the
question of the efficiency of gas ejection due to galactic winds remains open.

Tidal interaction may be responsible not only for mass exchange between
galaxies during collisions or close fly-bys and between galaxies and the IGM,
but also for changes in galactic morphology. According to the estimates of
\citet{Tutukov--2006ARep...50..439T}, every galaxy in a cluster experiences a
collision at least once during its lifetime. In these collisions, the galaxies
may merge, lose their gaseous components, or be disrupted completely. A new
galaxy may also form from gas lost in galaxy collisions.

The ram pressure of the IGM gas, evaporation of gas, and sweeping-out of dust
are less efficient galactic mass-loss mechanisms, though they influence the
chemical evolution of galaxies and enrichment of the IGM.

The essence of the statistical mechanism is that, in the case of an equilibrium
distribution of the stars in the gravitational potential of the galaxy, there
will always be stars with velocities exceeding the escape velocity. As these
stars leave the potential well, the system relaxes to a new equilibrium
state. However, the timescale for the statistical mechanism is very large ---
close to a hundred relaxation times \citep{Binney--1987gady.book.....B}, where
the latter is
\begin{equation}
  \tau_\mathrm{relax} \sim \frac{0.1 N}{\ln N}\,\tau_\mathrm{dyn}  \;,
\end{equation}
where $N$ is the number of stars in the system and $\tau_\mathrm{dyn}$ the
dynamical timescale of the system. Typical galactic dynamical timescales are
$\sim 10^7 \-- 10^8$~yr. Even for $\sim 10^6~M_\odot$ galaxies, the relaxation
time exceeds the Hubble time. Other mass-loss mechanisms rewiewed by
\citet{Binney--1987gady.book.....B} for collisionless stellar systems are even
less efficient.

It is commonly believed that most stars are born in associations (see however,
the paper by \citet{Elmegreen--1996ApJ...466..802E}). The lifetimes of OB
associations from birth to decay is short, of the order of several million
years. Typical velocities of stars acquired during the decay are of the order
of $10$~km$/$s, according to various studies
\citep{Gvaramadze--2008A&A...490.1071G} and observations
\citep{Gies--1987ApJS...64..545G}. Other estimates limit the velocity range to
$2\--8$~km$/$s \citep{Brown--1997MNRAS.285..479B}. The virial velocities in
low-mass galaxies can be several km$/$s
\citep{Karachentsev--2004AJ....127.2031K}, and the escape velocity lower than
$20$~km$/$s
\citep{Bovill--2009ApJ...693.1859B,Dijkstra--2004ApJ...601..666D}. In the case
of disk galaxies, the ordered motions of the galactic matter may facilitate the
ejection of stars. The aim of our current study is to estimate this effect and
observational manifestations in dwarf galaxies.

In Section 1, we compute the probability of ejection of stars from spheroidal
and disk galaxies. In Section 2, we present the results of modeling the
evolution of dwarf disk galaxies taking into account the ejection of
stars. Section 3 discusses our results.

\section{STELLAR EJECTION MECHANISM}

To escape its galaxy, the kinetic energy of a star must be sufficient to bring
about the work against the gravitational field
\begin{equation}
  \frac{(\vecb{v} + \vecb{u})^2}{2} \geqslant -\Phi  \;,
\end{equation}
where $\vecb{v}$ is the instantaneous velocity of the OB association,
$\vecb{u}$ the stellar velocity relative to the association,
and $\Phi$ the gravitational potential at the location of the association, with
the potential at infinity being zero. The dynamical timescales for associations
($\sim 10^8$~yr) exceed their lifetimes ($\sim 10^7$~yr) due to the low
densities of associations, which are $0.1~M_\odot/$pc$^3$
\citep{Brown--1997MNRAS.285..479B}. This means that energy equipartition for
stars of different masses does not have time to become established in the
association. Therefore, we will assume that the velocities of stars of any mass
have isotropic Gaussian distributions with dispersion $\sigma_\mathrm{OB}^2$;
the probability that a star moving away from the center of the OB association
will overcome the potential of its galaxy will then be
\begin{equation}
  \chi(\vecb{v}, -\Phi) =
  \int \limits_{\frac{(\vecb{v} + \vecb{u})^2}{2} \geqslant -\Phi}
  \frac{d^3u}{(2\pi\sigma_\mathrm{OB}^2)^{3/2}}\,
  \operatorname{exp}\left[-\frac{u^2}{2 \sigma_\mathrm{OB}^2}\right]  \;.
\end{equation}

The large-scale motion of the ISM in the galaxy increases the fraction of
ejected stars due to a sort of ``slingshot effect''. If the ISM participates in
the Keplerian motion of the galaxy with circular velocity $v$, the velocity of
a star after decay will be summed with the velocity of the association in the
galaxy. As a result, the probability of ejection of the star becomes (see the
Appendix)
\begin{equation}
  \label{eq:chi-keplerian}
  \chi(\eta, \psi) = 1 + \frac{1}{4} \int_0^\psi d\xi\,e^{-\xi/2} \left\{
  \operatorname{erf}\left[\frac{-\eta - \sqrt{\psi-\xi}}{\sqrt{2}}\right] -
  \operatorname{erf}\left[\frac{-\eta + \sqrt{\psi-\xi}}{\sqrt{2}}\right]
  \right\}  \;,
\end{equation}
where
\begin{equation}
  \eta = \frac{v}{\sigma_\mathrm{OB}} =
  \sqrt{\frac{r}{\sigma_\mathrm{OB}^2}\,\frac{\partial\Phi}{\partial r}}
  \;,\qquad
  \psi = - \frac{2 \Phi}{\sigma_\mathrm{OB}^2}  \;.
\end{equation}
Here, the velocity of the association is assumed to be equal to the local
velocity of the ISM. The reason for this is that the lifetimes of associations
are much shorter than the galactic dynamical timescale so the velocity of the
association does not change appreciably before the association decays. The
parameter $\eta$ in Eq. \ref{eq:chi-keplerian} describes the relative velocity
of the Keplerian motion in units of $\sigma_\mathrm{OB}$. Expression
\ref{eq:chi-keplerian} is also valid for galaxies in which large-scale motions
of gas are absent, such as spheroidal galaxies, but we must then adopt
$\eta = 0$.

The value of $\chi$ depends on the mass distribution in the galaxy and the
motion of the galactic gas. It is interesting to estimate the fraction of
ejected stars for some typical configurations of galaxies. We will calculate
this estimate as an average over the volume of the galaxy weighted by their
local star-formation rate $\Psi$:
\begin{equation}
  \label{eq:chi-mean}
  \overline{\chi} = \frac{\int_V dV\,\Psi\,\chi}{\int_V dV\,\Psi}  \;.
\end{equation}
Let us give estimates for a spheroidal galaxy with a Plummer density profile
and a disk galaxy with an exponential density profile.

\begin{figure*}[ht!]
  \centering
  \includegraphics[width=\textwidth]{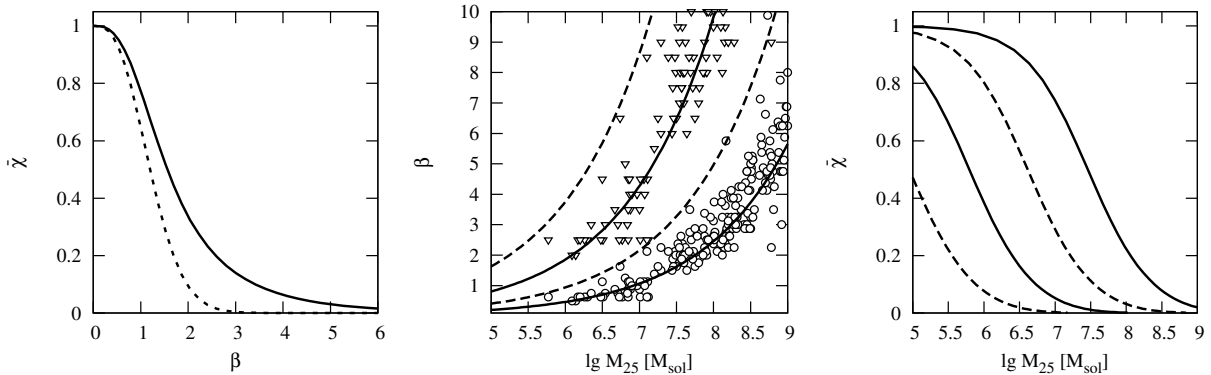}
  \caption{\footnotesize Left panel: average probability of ejection of stars
    from galaxies as a function of $\beta$, for disk galaxies with an
    exponential density profile and circular Keplerian rotation (solid) and
    non-rotating spheroidal galaxies with a Plummer density profile
    (dashed). The central panel shows $\beta$ as a function of the mass of the
    galaxy (up to $10^9~M_\odot$). The mass distribution for the galaxies was
    taken from the catalog of nearby galaxies of
    \citet{Karachentsev--2004AJ....127.2031K}. $M_{25}$ is the dynamical mass
    within the $25^m$ isophote. The parameter $\beta$ was computed as the ratio
    of the rotational velocity of neutral hydrogen (denoted in the catalog 
    $V_m$) and the velocity dispersion in the decaying OB association
    $\sigma_\mathrm{OB}$. Two limiting cases were considered for the latter:
    $\sigma_\mathrm{OB} = 2$~km$/$s (triangles) and
    $\sigma_\mathrm{OB} = 8$~km$/$s (circles). The solid lines show
    approximations for $\beta(M_{25})$ for these limiting cases. The dashed
    lines show analogous approximations for the ejection of stars outside the
    dark-matter halo (see the text). Right panel: dependence of the average
    probability of ejection of stars on the galaxy mass $M_{25}$. The lines
    correspond to the same cases as in central panel.}
  \label{fig:mass-beta-chi}
\end{figure*}

The distributions of the density and potential in an isotropic Plummer sphere
depend on two parameters --- the mass $M$ and the characteristic scale $a$:
\begin{equation}
  \rho = \frac{3 M}{4\pi a^3}\,(1 + r^2/a^2)^{-5/2}  \;,\qquad
  \Phi = - \frac{G M}{a}\,(1 + r^2/a^2)^{-1/2}  \,.
\end{equation}
Let us take the model of \citet{Firmani--1992A&A...264...37F} for the volume
star-formation rate:
\begin{equation}
  \Psi \propto \rho^2  \;.
\end{equation}
The distribution of the dimensionless potential $\psi$ over the scale $r/a$
depends only on the parameter $\beta$, which is defined as the ratio of the
typical virial velocity in the galaxy and the velocity dispersion in the
decaying association:
\begin{equation}
  \psi = \frac{2 \beta^2}{(1 + r^2/a^2)^{1/2}}  \;,\qquad
  \beta = \sqrt{\frac{G M}{\sigma_\mathrm{OB}^2 a}}  \;.
\end{equation}
As we noted above, we assume $\eta \equiv 0$ for a spheroidal galaxy. Inserting
these relations into Eq. \ref{eq:chi-mean} and integrating over the volume, we
obtain the coefficient $\overline{\chi}$ as a function of $\beta$
(Fig. \ref{fig:mass-beta-chi}, left panel, solid curve).

The distribution of the surface density of an exponential disk also depends on
the total mass and spatial scale:
\begin{equation}
  \Sigma = \frac{M}{2\pi a^2}\,e^{-r/a}  \;.
\end{equation}
Let us apply an approximate expression for the gravitational potential, with
the mass distribution taken to be spherically symmetric:
\begin{equation}
  \Phi = - \frac{G M}{a}\,\frac{1 - e^{-r/a}}{r/a}  \;.
\end{equation}
In this approximation, as in the case of a Plummer sphere, the distributions
for $\eta$ and $\psi$ depend only on $\beta$, which has the same definition:
\begin{equation}
  \eta = \beta\,\sqrt{\frac{1 - (1 + r/a)\,e^{-r/a}}{r/a}}  \;,\qquad
  \psi = 2 \beta^2\,\frac{1 - e^{-r/a}}{r/a}  \;.
\end{equation}
Finally, let us take the surface density of the starformation
rate according to the Schmidt–Kennicutt model
\citep{Kennicutt--1998ApJ...498..541K}:
\begin{equation}
  \Psi \propto \Sigma^{3/2}  \;.
\end{equation}

The function $\overline{\chi}(\beta)$ for the two cases is plotted in
Fig. \ref{fig:mass-beta-chi} (left). For disk galaxies, the values of $\beta$
for which the ejection of stars is possible are bounded from above by $4\--5$,
while this limit is $2\--2.5$ for spheroidal galaxies. We used data from the
Karachentsev catalog of nearby galaxies
\citep{Karachentsev--2004AJ....127.2031K} for galaxies with masses below
$10^9~M_\odot$ to derive the dependence of $\beta$ on the galactic mass. For
each galaxy with a given mass, the value of $\beta$ was calculated as the ratio
of the rotational velocity of neutral hydrogen (denoted in the catalog $V_m$)
and the velocity dispersion of stars in the decaying OB association
$\sigma_\mathrm{OB}$. Since $\sigma_\mathrm{OB}$ can be in the range
$2\--8$~km$/$s, the parameter $\beta$ is also not determined precisely. The
approximate mass dependence of this parameter is
\begin{equation}
  \label{eq:mass-beta}
  \beta = 0.025 \left(\frac{M_{25}}{M_\odot}\right)^{0.36}
  \left(\frac{\sigma_\mathrm{OB}}{\text{km$/$s}}\right)^{-1}  \;.
\end{equation}
In this approximation, we use the dynamical masses of the galaxies inside the
$25$ mag arcsec$^{-2}$ isophote, denoted $M_{25}$. Using only $M_{25}$ for the
mass underestimates the contribution of dark matter in the outer regions of the
galaxy but the approximation \ref{eq:mass-beta} remains valid even if we take
into account the dark haloes, viz. if we increase $M_{25}$ by factor equals to
$1 + \Omega_\mathrm{dm}/\Omega_\mathrm{b} \approx 6$, then the coefficient on
the right-hand side becomes $0.048$. The function \ref{eq:mass-beta} for the
limiting values of $\sigma_\mathrm{OB}$ is shown in
Fig. \ref{fig:mass-beta-chi} (middle panel) for the cases when the dark-matter
halo is and is not taken into account.

Below, we present results of our numerical modeling of dwarf galaxies in a
single-zone approximation, taking into account the ejection of stars. We traced
the chemical evolution of the galactic gas using the abundances of iron, oxygen
and heavy elements. We used the model of \citet{Firmani--1992A&A...264...37F}
and the studies of
\citet{Shustov--1997A&A...317..397S,Wiebe--1998ARep..42.....1W} as the basis
for the one-zone model.

\section{NUMERICAL MODELING}

\subsection{Single-zone Galactic Model}

In a decaying association, only stars with lifetimes $\lesssim~10^7$~yr end
their lives in the galaxy as Type II supernovae (SN II). These stars have
masses of about $13~M_\odot$ and higher
\citep{Tutukov--1980AZh....57..942T}. Some fraction of lower-mass stars, whose
number is determined by $\overline{\chi}$, leaves the galaxy and is not able to
participate in the enrichment of the ISM. However, this effect may be small,
since stars with masses below $8~M_\odot$ do not explode as SN II
\citep{Hashimoto--1993ApJ...414L.105H}, so that their disappearance influences
the chemical composition of the ISM only more weakly. On the other hand, Type
Ia supernovae (SN Ia) have delay times until their explosions of $10^8\--10^9$~
yr \citep{Tutukov--2002ARep...46..667T}, and their ejecta can strongly
influence the enrichment of the ISM in metals. The mass fraction of all stars
born as a single stellar population with a given initial mass function (IMF)
$\Phi(m)$ and leaving the galaxy is
$\overline{\chi} \int_{m_\mathrm{min}}^{13 M_\odot} dm\,\Phi$; for Salpeter IMF
with index $-2.35$ and stellar mass range $0.1~M_\odot\--100~M_\odot$ this
fraction is $0.88 \overline{\chi}$.

In the single-zone approximation, the total rate
of stellar mass loss due to the ejection of stars for
a galaxy with a time-dependent star formation rate
$\Psi(t)$ will be
\begin{equation}
  \dot{M}_\mathrm{s}^\mathrm{ej}(t)
  = \overline{\chi}\,\Psi(t - \tau_\mathrm{s}(13 M_\odot))
  \int_{m_\mathrm{min}}^{13 M_\odot} dm\,\Phi  \;,
\end{equation}
where $\tau_\mathrm{s} = \tau_\mathrm{s}(m)$ is the lifetime of a star with
mass $m$.

The ejection of stars also affects the amount of mass participating in star
formation and chemical enrichment of the ISM. The rate of gas return is
\begin{equation}
  \dot{M}_\mathrm{g}^\mathrm{fb}(t)
  = \int_{m_\mathrm{min}}^{m_\mathrm{max}} dm\,\Phi \Psi(t - \tau_\mathrm{s})
  \left(1 - \frac{m_\mathrm{r}}{m}\right)
  [ 1 - \overline{\chi}\,\theta(13 M_\odot - m) ]  \;,
\end{equation}
where $m_\mathrm{r} = m_\mathrm{r}(m)$ is the mass of the stellar remnant,
$\theta(x) = 0$ for $x \leqslant 0$, and $\theta(x) = 1$ for $x > 0$. The rate
of enrichment of the ISM in an element X is determined by the contributions
from SN I and SN II:
\begin{multline}
  \dot{M}_\mathrm{X}^\mathrm{fb}(t)
  = \frac{10^{-3}}{M_\odot}\,\Psi(t - \tau_\mathrm{SNIa})\,
  P_\mathrm{X}^\mathrm{SNIa}\,[1 - \overline{\chi}] + \\
  + \int_{m_\mathrm{min}}^{m_\mathrm{max}} dm\,\Phi \Psi(t - \tau_\mathrm{s})
  \left[ \left(1 - \frac{m_\mathrm{r}}{m}\right) X(t - \tau_\mathrm{s})
    + P_\mathrm{X}^\mathrm{SNII}(m, Z(t - \tau_\mathrm{s})) \right] \times \\
  \times [1 - \overline{\chi}\,\theta(13 M_\odot - m) ]  \;,
\end{multline}
where $\tau_\mathrm{SNIa}$ is the average delay time between the formation of a
binary and the SN Ia explosion, $P_\mathrm{X}^\mathrm{SNIa}$ is the mass of
element $\mathrm{X}$ ejected per explosion, and
$P_\mathrm{X}^\mathrm{SNII}(m, Z)$ is the mass of element $\mathrm{X}$ produced
by SN II with pre-supernova mass $m$ and initial heavyelement abundance $Z$.

In this model, we assumed that SN Ia are produced by mergers of degenerate
dwarfs in close binaries. As was shown by \citet{Tutukov--1994MNRAS.268..871T},
the mean delay time between the formation of a binary and the SN Ia explosion
is $\tau_\mathrm{SNIa} \approx 10^9$~yr. The rate of SN Ia explosions per unit
star-formation rate is obtained by normalizing to the modern SN Ia rate of
$0.003$ per yr and modern star formation rate of $3~M_\odot$ per yr
\citep{Tutukov--1994MNRAS.268..871T}. We assumed that each SN Ia produces
$P_\mathrm{Fe}^\mathrm{SNIa} = 0.6~M_\odot$ of iron
\citep{Tsujimoto--1995MNRAS.277..945T}; we did not consider the production of
other elements by these supernovae. The yields of SN II as functions of the
mass of the star and the initial heavy-element abundance were taken from
\citet{Maeder--1992A&A...264..105M}.

The balance of the total mass of galaxy, the mass of gas, and the masses of
various elements is determined not only by star formation, but also by
interaction of the galaxy with the ISM; i.e., via the galactic wind, dust
ejection, and the accretion of intergalactic gas. These factors were studied
previously for the same single-zone model (see, for instance, the study of
\citet{Shustov--1997A&A...317..397S}). However, accretion, dust ejection, and
galactic wind were not taken into account in the present study [the last factor
due to obvious shortcomings of the simple model for the galactic wind,
Eq. \ref{eq:galactic-wind}]. As a result, the mass balances of various
components of a galaxy was given by the equations:
\begin{equation}
  \label{eq:mass-balance}
  \begin{split}
    \dot{M}_\mathrm{tot}
    &= - \dot{M}_\mathrm{s}^\mathrm{ej}  \\
    \dot{M}_\mathrm{g}
    &= - \Psi + \dot{M}_\mathrm{g}^\mathrm{fb}  \\
    \dot{M}_\mathrm{X}
    &= - X \Psi + \dot{M}_\mathrm{X}^\mathrm{fb}  \;.
  \end{split}
\end{equation}
The details of this numerical model are presented in
\citet{Firmani--1992A&A...264...37F,Shustov--1997A&A...317..397S,
Wiebe--1998ARep..42.....1W}.

\subsection{Results of the Computations}

We can see from the dependence $\overline{\chi}(M_{25})$
(Fig. \ref{fig:mass-beta-chi}) that the decay of OB associations essentially
does not lead to mass loss from spheroidal galaxies, even those with the lowest
masses. For this reason, we restricted our numerical analysis to disk
galaxies.We computed four series of models for the evolution of galaxies with
masses from $10^{6.5}~M_\odot$ to $10^{8.5}~M_\odot$ and various values of
$\sigma_\mathrm{OB}$. For comparison, we also computed a series of closed
models. The radii of the galaxies corresponded to the relation $M \propto
R^2$. To avoid a large influence of the initial conditions on the burst of star
formation, the semi-thickness of the protogalactic disk was set to $10$~kpc in
all computations. The remaining parameters for the computed series of models
are given in the table. In series A, $\sigma_\mathrm{OB} = 2$~km$/$s, while in
series B and C $\sigma_\mathrm{OB} = 8$~km$/$s. In series C, we allowed for the
influence of the dark-matter halo [see the comments concerning
  Eq. \ref{eq:mass-beta}]. Thus, the series B computations enabled us to
estimate the effects of ejecting stars from the disk into the galactic halo,
while series C illustrated the effects of ejecting stars from the halo into the
IGM.
\begin{figure*}[ht!]
  \begin{center}
    \begin{tabular}{|c|c|c|c|c|c|c|c|}
      \hline
      \multicolumn{2}{|c|}{} &
      \multicolumn{2}{c|}{$
        \begin{array}{c}
          \sigma_\mathrm{OB} = 2~\text{km$/$s}
          \\ \text{(series A)}
        \end{array}
        $} &
      \multicolumn{2}{c|}{$
        \begin{array}{c}
          \sigma_\mathrm{OB} = 8~\text{km$/$s}
          \\ \text{(series B)}
        \end{array}
        $} &
      \multicolumn{2}{c|}{$
        \begin{array}{c}
          \sigma_\mathrm{OB} = 8~\text{km$/$s}
          \\ \text{+dark halo}
          \\ \text{(series C)}
        \end{array}
        $} \\
      \hline
      \makebox[2.0cm]{$\lg M/M_\odot$} & \makebox[1.5cm]{$R$, pc} &
      \makebox[0.5cm]{$\beta$} & \makebox[0.5cm]{$\overline{\chi}$} &
      \makebox[0.5cm]{$\beta$} & \makebox[0.5cm]{$\overline{\chi}$} &
      \makebox[0.5cm]{$\beta$} & \makebox[0.5cm]{$\overline{\chi}$} \\
      \hline
      $6.5$ & $79.5$ & $2.73$ & $0.21$ & $0.70$ & $0.92$ & $1.38$ & $0.6$ \\
      $7$ & $141.4$ & $4.14$ & $0.09$ & $1.07$ & $0.743$ & $2.08$ & $0.32$ \\
      $7.5$ & $251.5$ & $6.26$ & $0.012$ & $1.62$ & $0.481$ & $3.16$ & $0.13$ \\
      $8$ & $447$ & $9.48$ & $\sim 0$ & $2.45$ & $0.21$ & $4.78$ & $0.035$ \\
      $8.5$ & $795$ & $14.25$ & $\sim 0$ & $3.72$ & $0.075$ & $7.23$ & $\sim 0$ \\
      \hline
    \end{tabular}
  \end{center}
\end{figure*}

\begin{figure*}[ht!]
  \centering
  \includegraphics[width=\textwidth]{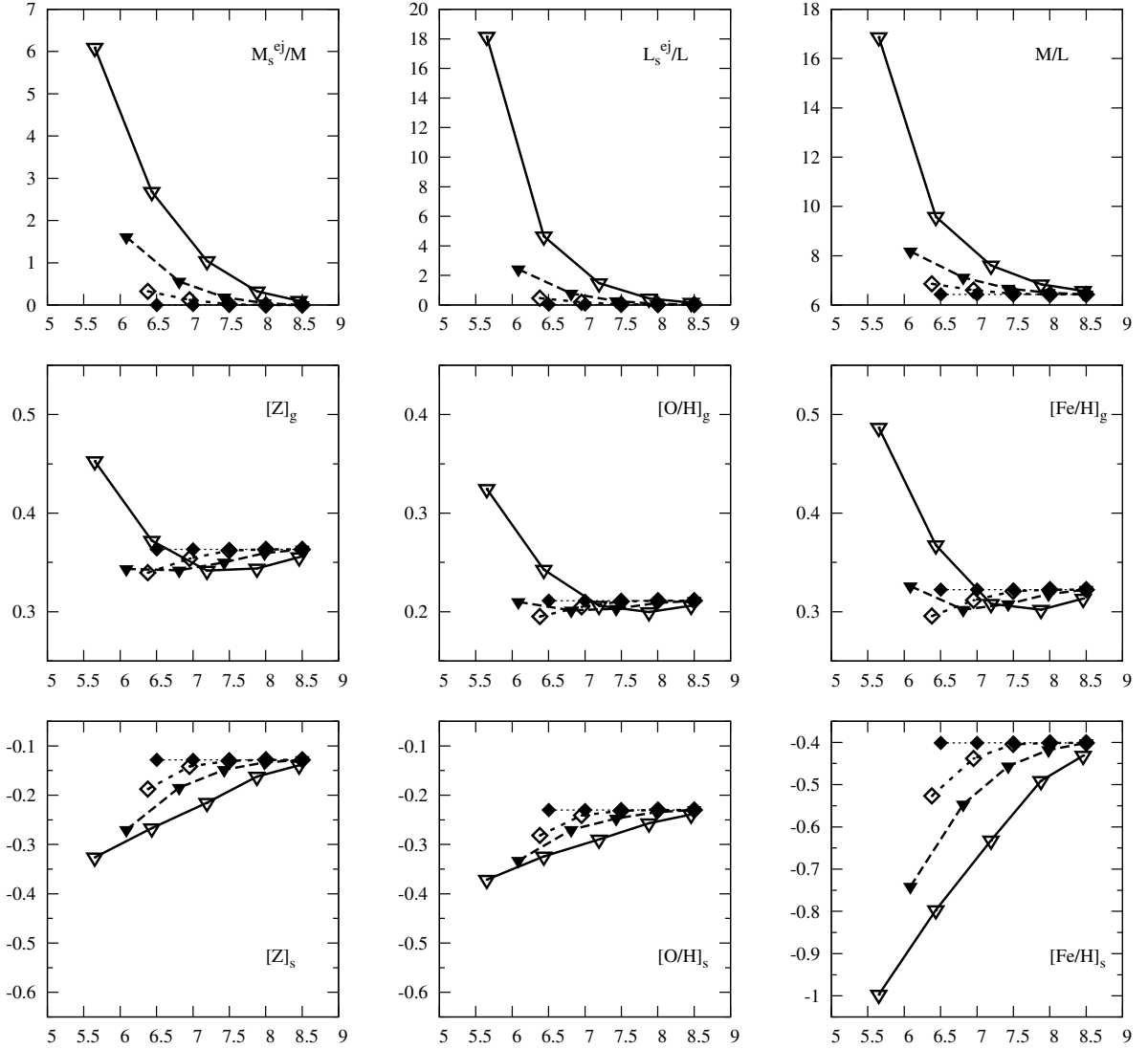}
  \caption{\footnotesize Dependence of integrated parameters on the disk masses
    (on a logarithmic scale in units of $M_\odot$) in four series of
    models. Solid diamonds correspond to the closed galaxy model, open diamonds
    to the model with $\sigma_\mathrm{OB} = 2$~km$/$s (series A), solid
    triangles to the model with $\sigma_\mathrm{OB} = 8$~km$/$s plus a
    dark-matter halo (series C), and open triangles to the model with
    $\sigma_\mathrm{OB} = 8$~km$/$s without a dark-matter halo (series B). The
    upper row of plots shows the ratio of the ejectedmass and diskmass, the
    ratio of the luminosities of the ejected stars and the disk, and the
    mass-to-luminosity ratio for the disk (in solar units). The panels in the
    middle and lower rows show the abundances of various elements in the gas
    and stars (averaged over the stellar population).}
  \label{fig:trends}
\end{figure*}

Figure \ref{fig:trends} shows the integrated characteristics of the model
galaxies at the end of the computations. As expected, including the ejection of
stars in the series A models did not result in significant deviations of the
integrated characteristics from those for the closed model. In the series B and
C models, for the lowest mass galaxies, the ratio of the ejected mass and the
dynamical mass can range from $1.5\--2$ (for matter ejected from the halo into
the IGM) to $6$ (for matter ejected from the disk into the halo). The amount of
ejected mass obtained in the series B models provides some idea of the extent
to which the morphology of a galactic disk can vary. For a galaxy with
dynamical mass $10^7~M_\odot$, the stellar mass in the halo exceeds the mass of
the disk by a factor of $1.5\--2$; i.e., such a galaxy cannot be considered a
disk galaxy. Galaxies with final dynamical masses of $10^6~M_\odot$ can have a
mass of $4.5 \times 10^6~M_\odot$ in the halo and $1.5 \times 10^6~M_\odot$ in
the IGM. The mass--luminosity relation shows that, compared to the closed model,
the total luminosity of the lowest-mass galaxies can be half an order of
magnitude lower, and the disk luminosity an order of magnitude lower. The
mass-to-luminosity ratio itself increases by more than a factor of $2.5$ for
the low-mass galaxies of series B. For series B galaxies with masses below
$10^7~M_\odot$, the total luminosity of stars ejected into the halo is a factor
of two to three higher than the disk luminosity (Fig. \ref{fig:trends}, middle
panel in the upper row). This also provides evidence for the transformation of
the galactic morphology from disk to spheroidal.

Since we did not consider galactic wind and accretion in our models, the
star-formation history in the galaxies consisted of a single burst which
depleted most of the gas: in all the galaxies, the mass fraction of the gas at
the end of the computations was $5\--10\%$. The relative abundances of elements
in the gas decrease weakly with the mass of the galaxy and with increasing
$\sigma_\mathrm{OB}$ (Fig. \ref{fig:trends}, middle row of plots). One
exception is the models at the low-mass end of series B, for which the relative
abundances of element grow with the galaxy mass by $0.1$ dex, compared to the
closed models. As the plots show, in all series of models, the mass dependence
of the abundances converges to a minimum for some mass in all series of
models. The reason for this may be a competition of two processes determining
the abundances in the gas: the enrichment of gas by supernovae and the return
of gas poor in elements by explosions of new low-mass stars. The influence of
the latter process may be diminished if a galaxy loses a considerable
fraction of its low-mass stars.

\begin{figure*}[ht!]
  \centering
  \includegraphics[angle=-90,width=\textwidth]{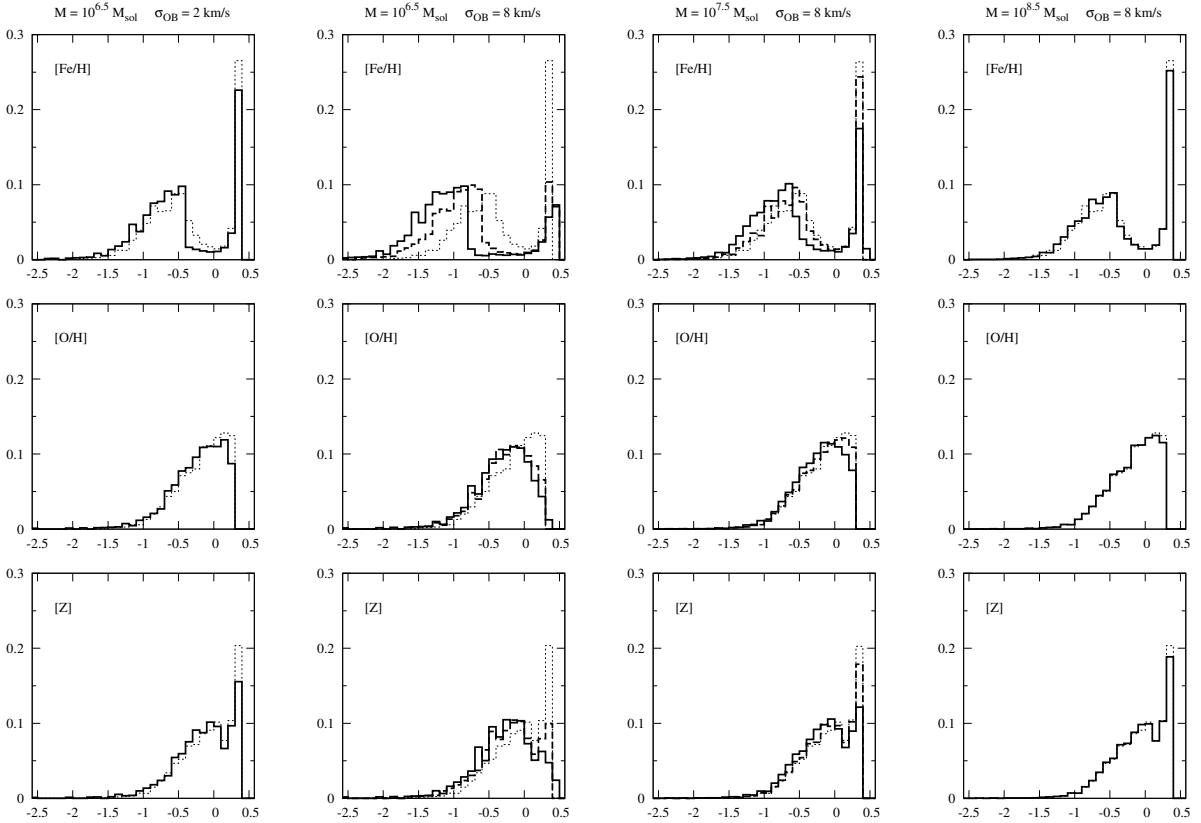}
  \caption{\footnotesize Distribution of the relative number of stars over the
    current abundances of iron, oxygen, and heavy elements. In the left column,
    the solid curves correspond to galaxies with masses of $10^{6.5} M_\odot$
    in the series A models, and the thin dotted curve to the closed model. In
    the other columns, the solid curves correspond to galaxies with masses of
    $10^{6.5} M_\odot$, $10^{7.5} M_\odot$, and $10^{8.5} M_\odot$ in the
    series B models, the dashed curves to galaxies with the same masses in the
    series C models, and the thin dotted curves to the closed model.}
  \label{fig:nstars}
\end{figure*}

The elemental abundances in the stars systematically decrease with the mass of
the galaxy and with increasing $\sigma_\mathrm{OB}$ (Fig. \ref{fig:trends},
lower row of plots). The ejection of stars influences the iron abundance
most. In a galaxy with dynamical mass $10^6~M_\odot$, the iron abundance can
decrease by $0.5$ dex in the disk stars (series B) and by more than $0.3$ dex
in the halo stars (series C). Figure \ref{fig:nstars} shows the distribution of
the relative numbers of stars over the abundances of iron, oxygen, and heavy
elements for all the computed series of models. Series B (second column of
plots) differs most from the closed model. Along with the shift of the
distribution toward lower abundances, stars heavily enriched in iron appear,
which formed in the first few billion years after the SN Ia explosions. The
characteristic appearance of the stellar distributions is due to the fact that
the galaxies in this model experience only one burst of star formation, which
occurred in a low-metallicity gas.

\section{DISCUSSION AND CONCLUSIONS}

We have studied the influence of the loss of stellar mass on the evolution of
dwarf spheroidal and disk galaxies. The decay of OB associations was considered
as a possible mass-loss mechanism, with the decay enabling some stars to obtain
velocities sufficient to escape their galaxy. The decay of associations is
essentially of no importance for the evolution of spheroidal galaxies. The
effect is also small for disk galaxies with
$\sigma_\mathrm{OB} = 2$~km$/$s. Since a value of $\sim 10$~km/s is thought to
be typical, we focused our analysis on models with
$\sigma_\mathrm{OB} = 8$~km/s. The results of our analysis are as follows.
\begin{enumerate}
  \item During the lifetime of an OB association ($\sim 10^7$~yr), the most
    massive SN II ($\gtrsim 13~M_\odot$) are able to enrich the ISM in the
    products of their explosions. Lower-mass stars that leave their galaxy do
    not contribute to the enrichment of the disk ISM, but instead serve as a
    source of elements for the halo or IGM. The same is true of SNIa.
  \item Disk galaxies that had at the onset of their star formation masses of
    $3 \times 10^7~M_\odot$ contain half of their mass in disk stars and the
    other half in the halo. The halo luminosities in such galaxies exceed the
    disk luminosities by a factor $1.5 \-– 2$. We can thus infer that galaxies
    with masses $\lesssim 10^7~M_\odot$ that were initially disk galaxies
    change their morphology to spheroidal. According to the classificaiton of
    \citet{de_Vaucouleurs--1991trcb.book.....D}, spiral galaxies are assigned
    morphological indices T = 4 (see also
    \citet{Corwin--1994AJ....108.2128C}). In the catalog of nearby galaxies
    \citep{Karachentsev--2004AJ....127.2031K}, disk galaxies have absolute
    magnitudes not exceeding $-13^m$ (for morphological indices from 0 to 7;
    i.e., including lenticular and irregular galaxies that are closest to disk
    galaxies); this magnitude corresponds to a luminosity $\sim 10^7~L_\odot$
    and a galactic mass $\sim 10^8~M_\odot$. Lower-mass galaxies are classified
    as spheroidal and irregular. This is confirmed by the computations for the
    models we have adopted here.
  \item In systems with masses $\lesssim 10^5~M_\odot$, a large fraction of the
    stellar mass leaves not only the disk of the galaxy, but also the halo
    (Fig. \ref{fig:mass-beta-chi}, right panel). Thus, if extremely low-mass
    galaxies can form at all, they can lose almost all their stellar population
    to the IGM after their first burst of star formation. As a result, a
    dark-matter halo enriched in gas should be left. This scenario may be
    important for the problem of missing satellites of the Galaxy.
  \item The ejection of stars increases the mass-toluminosity ratio. For
    galaxies with total masses (disk + halo) of $\sim 10^7~M_\odot$, this ratio
    increases by a factor of $2 \-- 2.5$ (Fig. \ref{fig:trends}, upper right
    panel).
  \item The ejection of stars may result in strong variations of elemental
    abundances in the gas (Fig. \ref{fig:trends},middle row of plots): along
    with the systematic decrease of the abundances by $\sim 0.05$ dex, the
    elemental abundances in the lowest-mass galaxies can increase by
    $0.1 \-- 0.15$ dex. The abundances in stars systematically fall with the
    galaxy mass decrease, by $0.2$ dex.
\end{enumerate}

This study was supported by the Federal Agency on Science and Innovation (state
contract no. 02.740.11.0247), the Federal Education Agency (contract
RNP-2.1.1-1937), the Program of State Support for Leading Scientific Schools of
the Russian Federation (grant no. NSh-4354.2008.2), and the Russian Foundation
for Basic Research (project nos. 08-02-91321-IND and 07-02-00454).

\section*{APPENDIX}

We will obtain an expression for the probability of ejection of a star from an
OB association moving along a Keplerian orbit in a galaxy with a given
gravitational potential $\Phi(r)$. A general expression for this probability is
\begin{equation}
  \label{eq:chi}
  \chi(\vecb{v}, -\Phi) =
  \int \limits_{\frac{(\vecb{v} + \vecb{u})^2}{2} \geqslant -\Phi}
  \frac{d^3u}{(2\pi\sigma_\mathrm{OB})^{3/2}}\,
  \operatorname{exp}\left[-\frac{u^2}{2 \sigma_\mathrm{OB}^2}\right]  \;.
\end{equation}
The velocity vector for a circular Keplerian orbit is
\begin{equation}
  \vecb{v} = \vecb{e}_\phi v  \;,\qquad
  v = \sqrt{r\,\frac{\partial\Phi}{\partial r}}  \;,
\end{equation}
where $\vecb{e}_\phi$ is a unit vector in the azimuthal direction. The
integration domain is in this case described by the equation
\begin{equation}
  u_\phi^2 + u_\perp^2 + 2 v u_\phi + v^2 + 2 \Phi \geqslant 0  \;,
\end{equation}
where $u_\perp$ is the modulus of the velocity orthogonal to
$\vecb{e}_\phi$. Solving this equation leads to the condition for the velocity
component $u_\phi$
\begin{equation}
  u_\phi \in (-\infty, \Re(u_{\phi-})] \cup [\Re(u_{\phi+}), +\infty)
  \;,\qquad
  u_{\phi\pm} = -v \pm \sqrt{- 2 \Phi - u_\perp^2}  \;.
\end{equation}
Since the components of the vector $\vecb{u}$ are distributed
isotropically and are independent, we can integrate
the general expression over $u_\phi$. Introducing the notation
\begin{equation}
  \xi = \frac{u_\perp^2}{\sigma_\mathrm{OB}^2}  \;,\qquad
  \eta = \frac{v}{\sigma_\mathrm{OB}}  \;,\qquad
  \psi = - \frac{2 \Phi}{\sigma_\mathrm{OB}^2}  \;,
\end{equation}
we finally obtain
\begin{equation}
  \chi(\eta, \psi) = 1 + \frac{1}{4} \int_0^\psi d\xi\,e^{-\xi/2} \left\{
  \operatorname{erf}\left[\frac{-\eta - \sqrt{\psi-\xi}}{\sqrt{2}}\right] -
  \operatorname{erf}\left[\frac{-\eta + \sqrt{\psi-\xi}}{\sqrt{2}}\right]
  \right\}  \;.
\end{equation}
If $\eta \equiv 0$, the general expression \ref{eq:chi} reduces to
\begin{equation}
  \chi(\psi)
  = \sqrt{\frac{2}{\pi}} \int_\psi^\infty d\zeta\,\zeta^2 e^{-\zeta^2/2}
  = 1 - \operatorname{erf}\left[\sqrt{\frac{\psi}{2}}\right]
  + \sqrt{\frac{2}{\pi}}\,\sqrt{\psi}\,e^{-\psi/2}  \;.
\end{equation}

\bibliography{preprint}
\bibliographystyle{plainnat}

\end{document}